\documentclass[submission,copyright,creativecommons]{eptcs}
\providecommand{\event}{EXPRESS/SOS 2023}

\usepackage{docmute}
\usepackage{graphicx}
\usepackage{color}
\usepackage{xcolor}
\usepackage{here}

\usepackage{amssymb}
\usepackage{amsfonts}
\usepackage{amsmath}
\usepackage{mathtools}
\usepackage{amsthm}
\usepackage{latexsym}
\usepackage{underscore}

\usepackage{tikz}
\usepackage{bussproofs}
\usepackage{stackengine}
\usepackage{listings}
\usepackage{wrapfig}
\usetikzlibrary{calc,decorations.pathmorphing,shapes,arrows}
\usepackage{subfig}
\usepackage{array}

\newcommand{\both}[2]{\xrightleftharpoons{#1}_{\hspace*{-2.6ex}\mathrm{#2}}}

\newcommand{\idp}{\mathrel{\iota}}
\newcommand{\lrhup}[2]{\ooalign{$#1\leftharpoondown$\cr$#1\rightharpoonup$\cr}}
\newcommand{\hpn}{\mathrel{\mathpalette\lrhup\relax}}

\newtheorem{prop}{Proposition}
\newtheorem{dfn}{Definition}
\newtheorem{lemm}{Lemma}
\newtheorem{theo}{Theorem}

\newcommand{\tran}[1]{\stackrel{#1}{\rightarrow}}
\newcommand{\ntran}[1]{\stackrel{#1}{\not\rightarrow}}
\newcommand{\dtran}[1]{\stackrel{#1}{\dashrightarrow}}

\newcommand{\anntran}[1]{\stackrel{#1}{\rightarrow}_{\mathrm{ann}}}

\newcommand{\Lab}{\mathsf{Lab}}
\newcommand{\ULab}{\underline{\mathsf{Lab}}}
\newcommand{\und}[1]{\mathsf{und}({#1})}
\newcommand{\idplab}{\mathrel{\iota}_{\mathrm{lab}}}

\newcommand{\semtran}{\xrightarrow{\texttt{sem}}}

\newcommand{\ltsi}{\mbox{\it LTSI}_{CRIL}}

\title{CRIL: A Concurrent Reversible Intermediate Language}
\author{Shunya Oguchi \qquad\qquad Shoji Yuen
\institute{Graduate School of Informatics, Nagoya University\\
Furo-cho, Chikusa-ku, Nagoya 464-8601, Japan}
\email{\{oguchi321,yuen\}@sqlab.jp}
}
\def\titlerunning{CRIL: A Concurrent Reversible Intermediate Language}
\def\authorrunning{S. Oguchi \&  S. Yuen}

\begin{document}
\maketitle

\begin{abstract}
    We present a reversible intermediate language with concurrency
    for translating a high-level concurrent programming language to
    another lower-level concurrent programming language, keeping reversibility.
    Intermediate languages are commonly used in compiling a source program
    to an object code program closer to the machine code, where an intermediate
    language enables behavioral analysis and optimization to be decomposed in steps.
    We propose CRIL (Concurrent Reversible Intermediate Language) as an extension
    of RIL used by Mogensen for a functional reversible language, incorporating a 
    multi-thread process invocation and the synchronization primitives based on the P-V
    operations. We show that the operational semantics of CRIL enjoy the
    properties of reversibility, including the causal safety and causal liveness proposed 
    by Lanese et al., checking the axiomatic properties.
    The operational semantics is defined by composing the bidirectional control flow
    with the dependency information on updating the memory, called {\em annotation DAG}.
    We show a simple example of `airline ticketing' to illustrate how CRIL
    preserves the causality for reversibility in imperative programs with concurrency.    
\end{abstract}

    \section{Introduction}

Reversible programming languages have been proposed to describe 
reversible computation where the control flows 
both forward and backward \cite{DBLP:conf/rc/YokoyamaAG11,DBLP:conf/rc/Hay-SchmidtGCH21,DBLP:journals/entcs/Yokoyama10,DBLP:journals/scp/HoeyU22}.
They directly describe reversible computation and develop new
aspects of software development since reversibility holds all information
at any point of execution.  In forward-only execution, the computation 
can overwrite the part of its intermediate history unless it is used
in the following computation for efficiency.
In analyzing the behavior, such as debugging, it is common to replay
the execution to the point in focus to recreate the lost part of history.
For a concurrent program, replaying the execution is usually difficult since
updating shared resources among multiple control threads 
depends on the runtime environment.

Intermediate languages mediate the translation from the source language 
to a low-level machine language for execution.  Step-by-step translation
via intermediate languages is a common technique for optimization in
compilers.  The intermediate language in LLVM \cite{DBLP:conf/cgo/LattnerA04} is often used as a behavioral
model for program analysis.

Mogensen uses RIL \cite{DBLP:conf/rc/Mogensen15} as an intermediate language with reversibility
for a functional reversible language in the memory usage analysis.
RSSA \cite{DBLP:conf/ershov/Mogensen15} based on RIL is used
for compiling and optimizing Janus programs \cite{DBLP:conf/rc/KutribMDS21,DBLP:conf/rc/DeworetzkiKMR22}.
Reversibility with concurrency has been studied in 
process calculi \cite{DBLP:conf/concur/DanosK04,DBLP:journals/jlp/PhillipsU07,DBLP:conf/concur/LaneseMS10,DBLP:journals/acta/LaneseMM21},
in event structures \cite{DBLP:journals/mscs/PhillipsU14,DBLP:journals/jlp/PhillipsU15,
DBLP:journals/ngc/UlidowskiPY18,DBLP:conf/lics/MelgrattiMP21}
and recently in programming languages such as Erlang \cite{DBLP:journals/jlp/LaneseNPV18} and
a simple imperative programming language \cite{DBLP:journals/scp/HoeyU22,DBLP:conf/rc/IkedaY20}.

We propose a reversible intermediate language CRIL by extending 
RIL.  CRIL extends RIL by allowing multiple blocks to run
concurrently and the synchronization primitive based on the P-V operations.
In CRIL, concurrent blocks interact with each other via shared variables.
To establish the reversibility for concurrent programs, the causality among
shared variables has to be preserved.   Unlike sequential reversible
programs, even if one step of a program is reversible, the whole program is not reversible in general 
since shared variables may not be reversed correctly.

To make a program of CRIL reversible, we give the operational semantics as the labeled
transition system, $\ltsi$, as
the composition of the operational semantics with one-step reversibility and
a data structure called `annotation DAG'.
An annotation DAG accumulates the causality of updating memory in a forward 
execution and rolls back the causality to control the
reversed flow in the backward execution.  We show that $\ltsi$ has the basic properties
for reversibility proposed in \cite{DBLP:conf/fossacs/LanesePU20}.  Using the approach of
\cite{DBLP:conf/fossacs/LanesePU20}, it is shown that $\ltsi$ enjoys the {\em Causal Safety}
and the {\em Causal Liveness}, which are important in analyzing CRIL programs compositionally.

By translating a high-level programming language to CRIL, $\ltsi$ works as a virtual machine,
and its behavior is guaranteed to be reversible.  CRIL enables fine-grained behavioral analysis
such as optimization and reversible debugging.  In section 4, we present a simple example of
airline ticketing given in \cite{DBLP:series/lncs/HoeyL0UV20} to enable reversible debugging.

The paper is organized as follows.  Section 2 presents the syntax of CRIL
and the operational semantics for control flow.  Section 3 introduces 
annotation DAG as a data structure to store the causality of updating
memory.  We define $\ltsi$ as the operational semantics for CRIL
and show the reversibility of $\ltsi$, which is followed by the airline ticketing example
in section 4.  Section 5 presents concluding remarks.

    \section{CRIL}

    \begin{wrapfigure}[13]{r}{6.5cm}
        \vspace*{-0.5cm}
        \centering
        \includegraphics{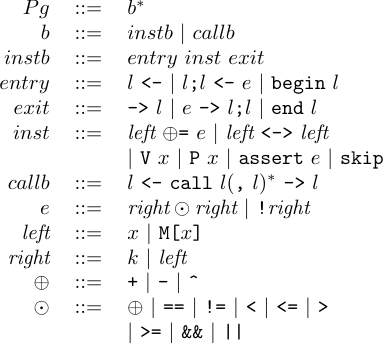}
        \caption{The syntax of CRIL}\label{fig:syntax}
    \end{wrapfigure}

    The syntax of CRIL is defined in figure~\ref{fig:syntax}.
    Following RIL \cite{DBLP:conf/rc/Mogensen15},
    a CRIL program consists of an unordered set of basic blocks.
    Given a set of labels $\mathcal{L}$, a block
    has an entry point followed by a block body and an exit point with labels.
    A block body is either a basic instruction or a call statement.

    \subsection{Basic block}

    We assume all references to variables have a global scope
    and there exists
    a heap memory $\texttt{M}$ indexed by integers, where
    $\texttt{M[}x\texttt{]}$ denotes the $x$-th element in $\texttt{M}$.
    An expression $e$ is either an arithmetic expression or a boolean
    expression with the usual operators
    $\texttt{+},\texttt{-},\texttt{\char`^},\texttt{==},
    \texttt{!=},\texttt{<},\texttt{<=},\texttt{>},\texttt{>=},\texttt{\&\&},
    \texttt{||},\texttt{!}$ of the C language, where
    $\texttt{\char`^}$ is the bitwise exclusive OR operation.
    The boolean operators and logical connectives treat $0$ as false and any non-$0$ value as true.
    An expression can contain integer constants, which are denoted by $k$.

    \paragraph{Entry/exit point}
    An entry/exit point of a basic block is the following forms:
    \begin{center}
        \begin{tabular}{ll|
        ll}
            \multicolumn{2}{c|}{Entry point} & \multicolumn{2}{c}{Exit point} \\ \hline
            (1)
            & $l\ \texttt{<-}$                   & (1') & $\texttt{->}\ l$ \\
            (2) & $l_1\texttt{;}l_2\ \texttt{<-}\ e$ & (2') & $e\ \texttt{->}\ l_1\texttt{;}l_2$   \\
            (3) & $\texttt{begin}\ l$                & (3') & $\texttt{end}\ l$
        \end{tabular}
    \end{center}
    where $l,l_1,l_2\in\mathcal{L}$. We write $\mathsf{entry}(b)$ for the entry point of
    a basic block $b$, and $\mathsf{exit}(b)$ for the exit point of a basic block $b$.

    The informal meaning of each item is explained as follows:
    \begin{trivlist}
        \item [(1) and (1'):] $l\ \texttt{<-}$ receives the control at $l$ unconditionally
        in a forward execution. In a backward execution, it sends the control to
        the block that receives the control at $l$.
        $\texttt{->}\ l$ dually works in the reversed way of $l\ \texttt{<-}$.
        \item [(2) and (2'):] $l_1;l_2\ \texttt{<-}\ e$ receives the control at $l_1$
        when $e$ is evaluated to a non-$0$ value and at $l_2$ otherwise in a forward
        execution. In a backward execution, it returns the control to the block
        that receives the control at $l_1$ when $e$ is evaluated to non-0 and
        at $l_2$ otherwise.   $e\ \texttt{->}\ l_1\texttt{;}l_2$ dually works
        in the reversed way of $l_1;l_2\ \texttt{<-}\ e$.
        \item [(3) and (3'):] $\texttt{begin}\ l$ receives the control from the call
        statement labeled by $l$ in a forward execution. In a backward execution,
        it returns the control to the statement labeled by $l$.
        $\texttt{end}\ l$ dually works in the reversed way of $\texttt{end}\ l$.
    \end{trivlist}

    A basic block is either an instruction block or a call statement.

    \paragraph{Instruction block}
    Basic instruction is in the forms:
    \begin{center}
        \begin{tabular}{llllll}
        (1)
            & ${\emph{left}}\oplus\!\texttt{=}\;e$      & (3) & $\texttt{V}\ x$ & (5) & $\texttt{assert}\ e$  \\
            (2) & ${\emph{left}}_1\ \texttt{{<->}}\ {\emph{left}}_2$ & (4) & $\texttt{P}\ x$ & (6) & $\texttt{skip}$
        \end{tabular}
    \end{center}
    We write $\textsf{inst}(b)$ for the basic instruction in $b$.
    The informal semantics is explained as follows:
    \begin{trivlist}
        \item [(1):] ${\emph{left}}\oplus\!\texttt{=}\;e$ is an {\em update} statement
        where ${\emph{left}}$ is a left-value, and
        $\oplus\in\{\texttt{+},\texttt{-},\texttt{\char`^}\}$.
        ${\emph{left}}$ is relatively updated by $e$ in that $\texttt{+=}$,
        $\texttt{-=}$, and $\texttt{\char`^=}$ with the same semantics
        as in the C language.
        If $\emph{left} = x$, $x$ must not appear in $e$.
        If $\emph{left} =$ \texttt{M[}$x$\texttt{]},
        heap references must not appear in $e$.
        \item [(2):] ${\emph{left}}_1\ \texttt{{<->}}\ {\emph{left}}_2$ is an {\em exchange} where
        ${\emph{left}}_1$ and ${\emph{left}}_2$ are left-values. It swaps the values
        specified by ${\emph{left}}_1$ and ${\emph{left}}_2$.
        The same variable must not appear on both sides of \texttt{{<->}}.
        \item [(3) and (4):] $\texttt{V}\ x$ and $\texttt{P}\ x$ are
        the P and V operations for synchronization, which correspond
        to those commonly used in operating systems. We assume variables
        in $\texttt{P}$ and $\texttt{V}$ instruction only appear as the parameters
        of $\texttt{P}$ and $\texttt{V}$.
        In a forward execution,
        $\texttt{V}\ x$
        is defined when $x$ is 0 and terminates when $x$ is 1 and
        $\texttt{P}\ x$ is defined when $x$ is $1$ and terminates
        when $x$ is $0$. In a backward execution, $\texttt{V}\ x$
        and $\texttt{P}\ x$ work
        as $\texttt{P}\ x$ and $\texttt{V}\ x$ of the forward execution
        respectively.
        \item [(5):] $\texttt{assert}\ e$ aborts the execution if $e$ evaluates
        to $0$, and does nothing otherwise.
        \item [(6):] $\texttt{skip}$ does nothing in either direction.
    \end{trivlist}

    We call $\mathcal{R} = Vars \cup\{\texttt{M}\}$ {\em memory resources}.
    Let $\mathsf{Var}(E)$ be the set of memory resource references appearing in $E$,
    where $E$ is one of $entry$, $exit$, or $inst$ in the grammar of figure~\ref{fig:syntax}.
    For example, $\mathsf{Var}(\texttt{z-=M[x]+y})=\{\texttt{M},\texttt{x},\texttt{y},\texttt{z}\}$.
    $\mathsf{read}(b)$ is the memory resources that $b$ uses, and
    $\mathsf{write}(b)$ is the memory resources that $b$ updates.
    \[\begin{array}{lll}
          \begin{array}{l}
              \mathsf{read}(b)=\mathsf{Var}(\mathsf{entry}(b)) \\
              \qquad\cup\;\mathsf{Var}(\mathsf{inst}(b))         \\
              \qquad\cup\;\mathsf{Var}(\mathsf{exit}(b))
          \end{array}
          & \quad &
          \mathsf{write}(b)=\begin{cases}
                                \{x\} & \mbox{If $\mathsf{inst}(b)=x\oplus\!\texttt{=}\;e$}\\
                                \{\texttt{M}\} & \mbox{If $\mathsf{inst}(b)=\texttt{M[}x\texttt{]}\oplus\!\texttt{=}\;e$}\\
                                \{x,y\} & \mbox{If $\mathsf{inst}(b)=x\;\texttt{{<->}}\;y$}\\
                                \{x,\texttt{M}\} & \mbox{If $\mathsf{inst}(b) \in \{x\;\texttt{{<->}}\;\texttt{M[}y\texttt{]},\texttt{M[}y\texttt{]}\;\texttt{{<->}}\;x\}$}\\
                                \{\texttt{M}\} & \mbox{If $\mathsf{inst}(b)=\texttt{M[}x\texttt{]}\;\texttt{{<->}}\;\texttt{M[}y\texttt{]}$}\\
                                \{x\} & \mbox{If $\mathsf{inst}(b)\in\{\texttt{P}\ x,\texttt{V}\ x\}$}\\
                                \varnothing          & \mbox{Otherwise.}
          \end{cases}
    \end{array}
    \]

    \paragraph{Call statement}

    A {\em call statement} is a basic block in the following form:
    \begin{center}
        $l\;\texttt{<-}\ \texttt{call}\ l_1\texttt{,}\cdots\texttt{,}l_n\ \texttt{->}\;l'$ ($n\geq 1$)
    \end{center}

    When $n=1$, it behaves as a subroutine call in RIL.
    If $n\geq 2$, the controls are simultaneously sent to all
    basic blocks with $\texttt{begin}\ l_i$ in the
    forward execution, and to all basic blocks with $\texttt{end}\ l_i$
    in the backward execution.
    In a forward execution, $\texttt{call}\ l_1\texttt{,}\cdots\texttt{,}l_n$ terminates
    when all controls are returned to this block. In a backward execution,
    it sends the controls to the blocks whose exit points are $\texttt{end}\ l_i$.

    As in RIL,
    $\texttt{call}$ appears only in a basic block whose entry and exit parts
    are unconditional, and not in any $\texttt{begin}$ and $\texttt{end}$
    blocks.
    CRIL does not have $\texttt{uncall}$ for a call statement
    since \texttt{uncall} makes the semantics more complex
    in that an uncall nested in multiple calls causes
    the mixture of forward and backward execution for process blocks.
    An uncall can be implemented as a symmetrical call.

    \subsection{Process}

    For a basic block $b$, $\mathsf{in}(b)$,$\mathsf{out}(b)\subseteq\mathcal{L}$ are defined as follows:

    \[
        \begin{array}{ll}
            \mathsf{in}(b)=\begin{cases}
                               \{l\} & \mbox{if}\ \mathsf{entry}(b)=l\;\texttt{<-}\\
                               \{l_1,l_2\} & \mbox{if}\ \mathsf{entry}(b)=l_1;l_2\;\texttt{<-}\;e\\
                               \varnothing & \mbox{if}\ \mathsf{entry}(b)=\texttt{begin}\ l
            \end{cases}
            &
            \mathsf{out}(b)=\begin{cases}
                                \{l\} & \mbox{if}\ \mathsf{exit}(b)=\texttt{->}\;l\\
                                \{l_1,l_2\} & \mbox{if}\ \mathsf{exit}(b)=e\;\texttt{->}\;l_1;l_2\\
                                \varnothing & \mbox{if}\ \mathsf{exit}(b)=\texttt{end}\ l
            \end{cases}
        \end{array}
    \]

    Basic blocks $b_1$ and $b_2$ are connected, written as $b_1\bowtie b_2$, if $\mathsf{out}(b_1)\cap\mathsf{in}(b_2)\not=\varnothing$
    or $\mathsf{in}(b_1)\cap\mathsf{out}(b_2)\not=\varnothing$.
    A {\em process block} of $b$ is $\mathsf{PB}(b,Pg)=\{b'\in Pg\;|\;b'\bowtie^\ast b\}$,
    where $\ast$ stands for reflexive and transitive closure.
    No basic block is shared among process blocks since
    they are the equivalence classes of $\bowtie^*$,
    which is an equivalence relation on basic blocks.

    Let $L_1(B) = \bigcup_{b\in B} (\mathsf{in}(b)\cup\mathsf{out}(b))$
    and $L_2(B) = \{l \;|\; \exists b\in B.\;
    \mathsf{entry}(b) = \texttt{begin}\;l \lor \mathsf{exit}(b) = \texttt{end}\;l\}$.
    A CRIL program $Pg$ is {\em well-formed} when it satisfies the following conditions:
    \begin{trivlist}
        \item [(1):] For all $l\in L_1(Pg)$, there exists a unique pair $(b_1,b_2)$
        such that $\mathsf{in}(b_1) \cap \mathsf{out}(b_2) = \{l\}$;
        \item [(2):] For all $l\in L_2(Pg)$, there exists a unique pair $(b_1,b_2)$
        such that $\mathsf{entry}(b_1)=\texttt{begin}\;l$ and
        $\mathsf{exit}(b_2)=\texttt{end}\;l$;
        \item [(3):] $L_1(Pg) \cap L_2(Pg) = \varnothing$
        \item [(4):] For all $b\in Pg$, $|L_2(\mathsf{PB}(b,Pg))|=1$; and
        \item [(5):] There is a special label $\texttt{main}\in L_2(Pg)$.
    \end{trivlist}

    The well-formedness ensures that once a control enters into a process block at $\texttt{begin}\ l$,
    the control may reach only the matching $\texttt{end}\ l$ in the forward execution and vice versa in the
    backward execution.
    A process block $\mathsf{PB}(b,Pg)$ is {\em labeled} by $l$
    when it contains the basic block with $\texttt{begin}\ l$.

    A call statement $\texttt{call}\ l_1\texttt{,}\cdots\texttt{,}l_n$ sends controls to process blocks labeled by
    $l_1,\cdots,l_n$. A process block with control is called a {\em process}.
    A process is executed by passing the control among the basic blocks in its process block.
    Since $\texttt{call}$ can be recursive, a process may have some subprocesses.

    In a forward execution, $l\ \texttt{<-}\ \texttt{call}\ l_1\texttt{,}\cdots\texttt{,}l_n\ \texttt{->}\ l'$
    receives a control at $l$, forks $n$ processes, and sends the control to $l'$ after merging
    the processes. In the backward execution, it works in a reversed manner.
    In the following, we assume a CRIL program is well-formed.

    \subsection{Basic operational semantics}

    The set of process identifiers $\mathsf{PID}$ is
    $(\mathbb{N}_+)^\ast$ where ${\mathbb{N}_+}$ is the set of positive integers.
    $p\in\mathsf{PID}$ denotes an identifier uniquely assigned to
    a process.
    When $p$ executes a process block $\mathsf{PB}(b,Pg)$, we also write
    $\mathsf{PB}(p)$. If $p$ is labeled by $l$, $\mathsf{PB}(p)=
    \mathsf{PB}(b,Pg)$ where $\mathsf{entry}(b)=\texttt{begin}\ l$.
    A special {\em root} process has the identifier of $\varepsilon$. The
    runtime invokes the root process and sends the control to a process block
    labeled by $\texttt{main}$ to start an execution
    of a CRIL program. For a process $p$, $p\cdot i$
    is assigned to the $i$-th subprocess invoked by a call statement
    of process $p$.
    $\preceq$ is the prefix relation.
    A process set $PS$ is a set of process identifiers
    satisfying (1) $\varepsilon\in PS$; (2) $p\in PS$ implies $p'\in PS$ for $p'\preceq p$;
    and (3) $p\cdot i$ implies $p\cdot j\in PS$ for $j<i$.
    For a process set $PS$ and a process id $p$, $\mathsf{isleaf}(PS,p)$
    holds if for all $p'\in PS$, $p\preceq p'$ imples $p=p'$.

    A {\em process configuration} is $(l, stage)$,
    where $l \in \mathcal{L}$ and $stage \in \{\texttt{begin}, \texttt{run}, \texttt{end}\}$
    are the location of the control in a process block.
    If $stage=\texttt{begin}$, it is before executing the process block,
    if $stage=\texttt{run}$, it is executing the process block, and
    if $stage=\texttt{end}$ it terminated the process block.
    $\mathsf{PC}$ is the set of process configurations.

    A {\em program configuration} is $({ Pg }, \rho, \sigma, { Pr })$, where
    ${ Pg }$ is the program (which never changes),
    $\rho : Vars \to \mathbb{Z}$ maps a variable to its value,
    $\sigma : \mathbb{N} \to \mathbb{Z}$ maps a heap memory address to its value.
    A {\em process map}
    ${ Pr } : \mathsf{PID} \to \mathsf{PC}\cup\{\bot\}$ maps a process
    to a process configuration.
    We assume $Pr_{act}$ is a process set
    where $Pr_{act}=\{p\in\mathsf{PID} | Pr(p)\in\mathsf{PC}\}$.
    $\mathcal{C}$ is the set of all program configurations.

\begin{figure}[H]
\begin{center}
    \hspace*{-0.18cm}
\includegraphics[scale=0.855]{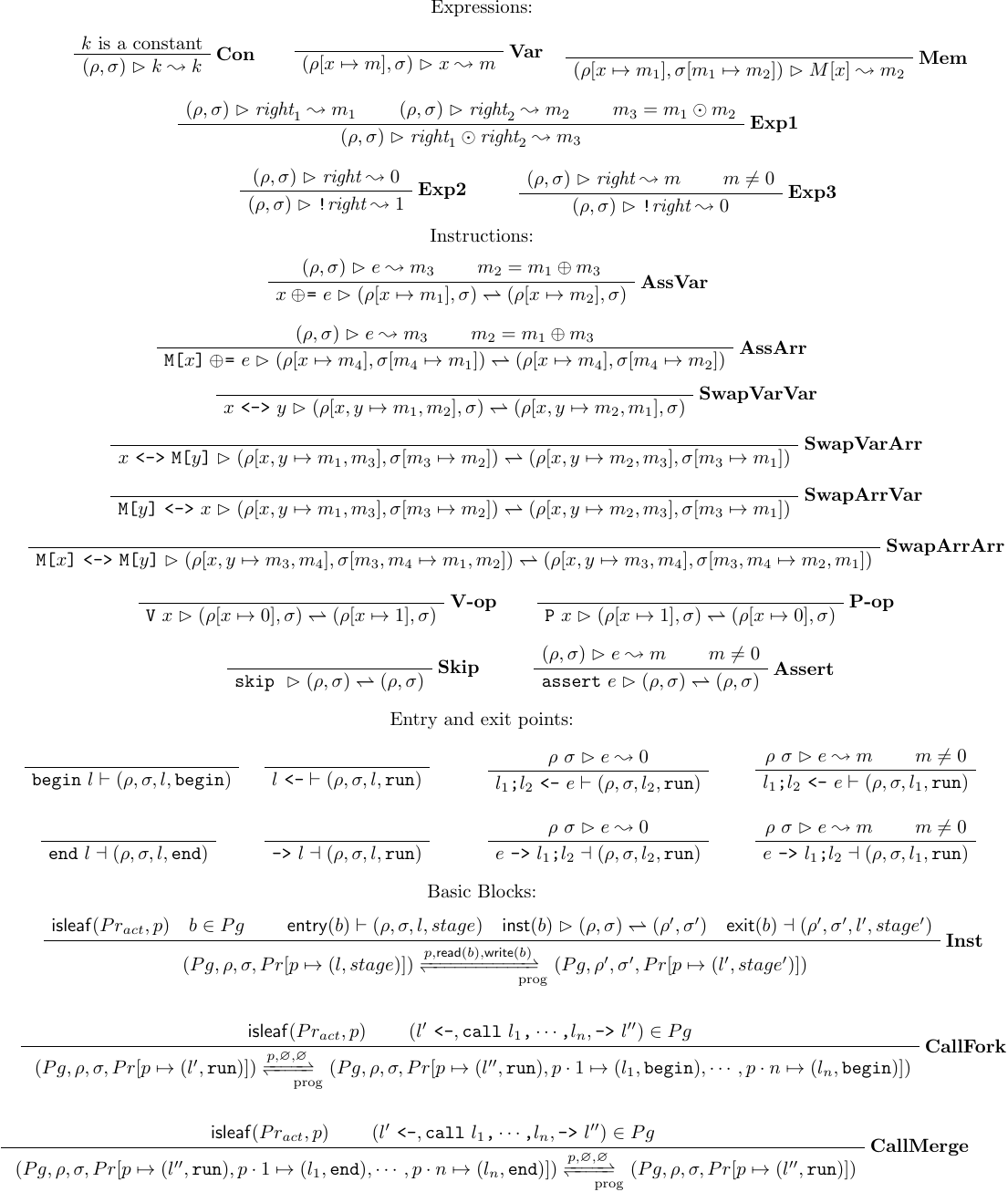}
\end{center}

\caption{The basic operational semantics}\label{SOS1}
\end{figure}

    A transition relation over program configurations
    \[(Pg,\rho,\sigma,Pr)\both{p,Rd,Wt}{prog}(Pg,\rho',\sigma',Pr')\]
    is defined in figure~\ref{SOS1}.
    $(Pg,\rho,\sigma,Pr)$ steps forward to $(Pg,\rho',\sigma',Pr')$
    by the process $p$ with reading memory resource $Rd$ and
    updating memory resource $Wt$. And $(Pg,\rho',\sigma',Pr')$
    steps backward to $(Pg,\rho,\sigma,Pr)$ in the same way.

    We explain the SOS rules in figure~\ref{SOS1}.
        {\bf AssVar} and {\bf AssArr} present the update behavior.
    The exchange behavior is presented by
        {\bf SwapVarVar}, {\bf SwapVarArr}, {\bf SwapArrVar}, and {\bf SwapArrArr}.
        {\bf SwapVarArr} and {\bf SwapArrVar} are reversible
    since $y$ is evaluated to the same value on both sides of $\hpn$.
    {\bf SwapVarVar} and {\bf SwapArrArr} are clearly reversible.
        {\bf Skip} presents the skip behavior.
        {\bf Assert} presents the assertion behavior,
    which stops when $e$ is evaluated to $0$.

        {\bf V-op} and {\bf P-op} present the behavior of \texttt{V}\ $x$
    and \texttt{P}\ $x$
    for synchronization by $x$ shared among concurrent processes.
    In forward execution, \texttt{V}\ $x$ sets $x=1$ when $x=0$, and waits otherwise.
    In backward execution, \texttt{V}\ $x$ sets $x=0$ when $x=1$, and waits otherwise.
    \texttt{P} behaves in a symmetrical fashion.
    By the pair of \texttt{V}\ $x$ and \texttt{P}\ $x$,
    $x$ can be used as a semaphore to implement the mutual exclusion
    for both directions of execution.

    {\bf Inst} presents the one-step behavior of
    a basic block. The instruction updates $\rho$ and $\sigma$
    and the entry and exit points give the status of the process.
    The process is running if $stage$ is $\texttt{run}$.
    the process is at the initial
    block or at the final block, if $stage$
    is $\texttt{begin}$ or $\texttt{end}$.
    The transition label $Rd$ is
    $\texttt{read}(b)$ and the transition label
    $Wt$ is $\texttt{write}(b)$.

    {\bf CallFork} presents that a call statement forks subprocesses.
    When $p$ executes a call statement $\texttt{call}\ l_1\texttt{,}\cdots\texttt{,}l_n$ forwards,
    it forks subprocesses labeled by $l_1,\cdots,l_n$ and $p$ stores the label
    for returning the controls in $Pr$.
    Note that the process map is changed to $Pr'$ with subprocesses after forking subprocesses.
    Since $\mathsf{isleaf}(Pr'_{act},p)$ does not hold, $p$ does not pass the control
    to the next block until all the subprocesses are merged.
        {\bf CallMerge} works dually to {\bf CallFork}. In a forward execution,
    when all subprocesses reach the ${\texttt{end}}$ stage, all subprocesses are
    set to inactive and $p$ resumes to pass the control to the next basic block.
    In a backward execution, ${\bf CallFork}$ behaves as ${\bf CallMerge}$ of forward
    execution and vice versa for ${\bf CallMerge}$.

\newpage

    In a program configuration of CRIL, there is no stack as in RIL to store the return label
    for subroutine calls. The process map stores the return label, which is not available
    until $\mathsf{isleaf}(Pr_{act},p)$ holds, where it checks if the label is on the
    stack.

    Figure\ref{lst:shared} shows an example of CRIL program $Pg$.
    There are four process blocks $\{b_1,b_2,b_3\}$,$\{b_4,b_5\}$,
    $\{b_6\}$, and $\{b_7\}$. A process map assigns $\varepsilon$ to
    $\{b_1,b_2,b_3\}$. In the following execution, it assigns
    1 to $\{b_4,b_5\}$, 2 to $\{b_6\}$, and 3 to $\{b_7\}$.

    An example of the transitions for $Pg$ is as follows:

\begin{wrapfigure}[5]{R}[0pt]{0.4\textwidth}
\vspace*{-5.2cm}
\includegraphics[width=0.4\textwidth]{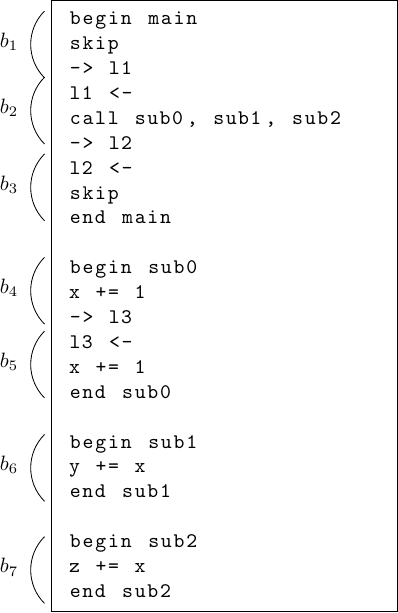}
\caption{A CRIL program $Pg$}\label{lst:shared}
\end{wrapfigure}

    \noindent
    {\small
        $\begin{array}{l}
        (Pg,\rho_0,\sigma_0,[\varepsilon\mapsto(\texttt{main},\texttt{begin})])
             \\
             \both{\varepsilon,\varnothing,\varnothing}{prog}
             (Pg,\rho_0,\sigma_0,[\varepsilon\mapsto(\texttt{l1},\texttt{run})])\\
             \both{\varepsilon,\varnothing,\varnothing}{prog}
             (Pg,\rho_0,\sigma_0,\begin{array}{l}
                                     [\varepsilon\mapsto(\texttt{l2},\texttt{run}),1\mapsto(\texttt{begin},\texttt{sub0}),\\
                                     2\mapsto(\texttt{sub1},\texttt{begin}),3\mapsto(\texttt{sub2},\texttt{begin})]
             \end{array})\\
             \both{1,\{\texttt{x}\},\{\texttt{x}\}}{prog}
             (Pg,\rho_1,\sigma_0,\begin{array}{l}
                                     [\varepsilon\mapsto(\texttt{l2},\texttt{run}),1\mapsto(\texttt{l3},\texttt{run}),\\
                                     2\mapsto(\texttt{sub1},\texttt{begin}),3\mapsto(\texttt{sub2},\texttt{begin})]
             \end{array}
             )\\
             \qquad \mbox{where}\ \rho_1=\rho_0[\texttt{x}\mapsto 1]\\
             \both{2,\{\texttt{x},\texttt{y}\},\{\texttt{x}\}}{prog}
             (Pg,\rho_2,\sigma_0,\begin{array}{l}
                                     [\varepsilon\mapsto(\texttt{l2},\texttt{run}),1\mapsto(\texttt{l3},\texttt{run}),\\
                                     2\mapsto(\texttt{sub1},\texttt{end}),3\mapsto(\texttt{sub2},\texttt{begin})]
             \end{array}
             )\\
             \qquad \mbox{where}\ \rho_2=\rho_2[\texttt{y}\mapsto 1]\\
             \both{3,\{\texttt{x},\texttt{z}\},\{\texttt{z}\}}{prog}
             (Pg,\rho_3,\sigma_0,\begin{array}{l}
                                     [\varepsilon\mapsto(\texttt{l2},\texttt{run}),1\mapsto(\texttt{l3},\texttt{run}),\\
                                     2\mapsto(\texttt{sub1},\texttt{end}),3\mapsto(\texttt{sub2},\texttt{end})]
             \end{array}
             )\\
             \qquad \mbox{where}\ \rho_3=\rho_1[\texttt{z}\mapsto 1]\\
             \both{1,\{\texttt{x}\},\{\texttt{x}\}}{prog}
             (Pg,\rho_4,\sigma_0,\begin{array}{l}
                                     [\varepsilon\mapsto(\texttt{l2},\texttt{run}),1\mapsto(\texttt{sub0},\texttt{end}),\\
                                     2\mapsto(\texttt{sub1},\texttt{end}),3\mapsto(\texttt{sub2},\texttt{end})]
             \end{array}
             )\\
             \qquad \mbox{where}\ \rho_4=\rho_3[\texttt{x}\mapsto 2]\\
             \both{\varepsilon,\varnothing,\varnothing}{prog}
             (Pg,\rho_4,\sigma_0,
             [\varepsilon\mapsto(\texttt{l2},\texttt{run})])\\
             \both{\varepsilon,\varnothing,\varnothing}{prog}
             (Pg,\rho_4,\sigma_0,
             [\varepsilon\mapsto(\texttt{main},\texttt{end})])\\
        \end{array}$
    }

    This forward execution ends with $\texttt{x}=2,\texttt{y}=1,\texttt{z}=1$.
    The operational semantics show that the computation may be reversed to
    $(Pg,\rho_0,\sigma_0,[\varepsilon\mapsto(\texttt{main},\texttt{begin})])$.
    However, it is possible to reverse to a different configuration
    such as $\texttt{x}=0,\texttt{y}=-1,\texttt{z}=-1$ if the call statement is reversed in a different
    order. Thus, this operational semantics is not reversible. In the next
    section, we will combine an annotation for the dependency information
    as DAG to make the basic properties for
    reversibility as well as Causal Safety and Causal Liveness.

    \section{Reversibility of CRIL}

    Table~\ref{tbl:exec}\;(a) shows the transitions of store $\rho$
    by the sequence of basic blocks in the forward computation
    of the example in the previous section.
    Process $p$ makes the forward (left-to-right) transition of
    $\both{p,Rd,Wt}{prog}$.
    The program configuration at the end is
    $(Pg,[\texttt{x}\mapsto 2,\texttt{y}\mapsto 1,\texttt{z}\mapsto 1],\sigma_0,
    [\varepsilon\mapsto(\texttt{main},\texttt{end})]$.
    The configuration may lead to a different store by
    the backward (right-to-left) transitions of $\both{p,Rd,Wt}{prog}$
    as shown in table~\ref{tbl:exec}\;(b). Although each step of the
    operational semantics keeps the local reversibility, it does not
    preserve the causality of shared memory. The
    forward step of $\both{p,Rd,Wt}{prog}$ updates $Wt$ reading
    $Rd$ making the causality from $Rd$ to $Wt$.
    Our idea is to control processes to keep the causality
    by observing $Rd$ and $Wt$
    being combined with the operational semantics.

    \begin{minipage}[t]{0.55\textwidth}
        \begin{table}[H]
            \vspace*{-0.5cm}
            \hspace*{-0.75cm}
            \subfloat[][A forward Execution]{\small
                \begin{tabular}{|l|c|c|c|}
                    \hline
                    & \texttt{x} & \texttt{y} & \texttt{z} \\ \hline
                    & 0          & 0          & 0          \\
                    $b_1\in \mathsf{PB}(\varepsilon)$ &            &            &            \\
                    & 0          & 0          & 0          \\
                    $b_2\in \mathsf{PB}(\varepsilon)$ &            &            &            \\
                    & 0          & 0          & 0          \\
                    $b_4\in \mathsf{PB}(1)$ &            &            &            \\
                    & 1          & 0          & 0          \\
                    $b_6\in \mathsf{PB}(2)$ &            &            &            \\
                    & 1          & 1          & 0          \\
                    $b_7\in \mathsf{PB}(3)$ &            &            &            \\
                    & 1          & 1          & 1          \\
                    $b_5\in \mathsf{PB}(1)$ &            &            &            \\
                    & 2          & 1          & 1          \\
                    $b_2\in \mathsf{PB}(\varepsilon)$ &            &            &            \\
                    & 2          & 1          & 1          \\
                    $b_3\in \mathsf{PB}(\varepsilon)$ &            &            &            \\
                    & 2          & 1          & 1          \\
                    \hline
                \end{tabular}
            }\quad
            \subfloat[][The corresponding backward execution]{\small
                \begin{tabular}{|l|c|c|c|}
                    \hline
                    & \texttt{x} & \texttt{y} & \texttt{z} \\ \hline
                    & 2          & 1          & 1          \\
                    $b_3\in \mathsf{PB}(\varepsilon)$ &            &            &            \\
                    & 2          & 1          & 1          \\
                    $b_2\in \mathsf{PB}(\varepsilon)$ &            &            &            \\
                    & 2          & 1          & 1          \\
                    $b_7\in \mathsf{PB}(3)$ &            &            &            \\
                    & 2          & $-1$         & 1          \\
                    $b_6\in \mathsf{PB}(2)$ &            &            &            \\
                    & 2          & $-1$         & $-1$         \\
                    $b_5\in \mathsf{PB}(1)$ &            &            &            \\
                    & 1          & $-1$         & $-1$         \\
                    $b_4\in \mathsf{PB}(1)$ &            &            &            \\
                    & 0          & $-1$         & $-1$         \\
                    $b_2\in \mathsf{PB}(\varepsilon)$ &            &            &            \\
                    & 0          & $-1$         & $-1$         \\
                    $b_1\in \mathsf{PB}(\varepsilon)$ &            &            &            \\
                    & 0          & $-1$         & $-1$         \\
                    \hline
                \end{tabular}
            }
            \caption[]{Store changes in executions}\label{tbl:exec}
        \end{table}
    \end{minipage}
    \begin{minipage}[t]{0.4\textwidth}
        \begin{table}[H]
            \centering
            \begin{tabular}{|c|c|c|}\hline
            & $\mathsf{read}$             & $\mathsf{write}$ \\ \hline
            $b_1$ & $\varnothing$ & $\varnothing$    \\
            $b_2$ & $\varnothing$               & $\varnothing$    \\
            $b_3$ & $\varnothing$               & $\varnothing$    \\
            $b_4$ & $\{\mathtt{x}\}$            & $\{\mathtt{x}\}$ \\
            $b_5$ & $\{\mathtt{x}\}$            & $\{\mathtt{x}\}$ \\
            $b_6$ & $\{\mathtt{x},\mathtt{y}\}$ & $\{\mathtt{y}\}$ \\
            $b_7$ & $\{\mathtt{x},\mathtt{z}\}$ & $\{\mathtt{z}\}$ \\ \hline
            \end{tabular}
            \caption{$\mathsf{read}$ and $\mathsf{write}$ for basic blocks}
            \label{tbl:ReadWrite-example}
        \end{table}

        \quad Table~\ref{tbl:ReadWrite-example} presents $\mathsf{read}$
        and $\mathsf{write}$ for each basic block.
        In the backward execution, after reversing $b_3b_2$,
            {\bf CallMerge} works as a forking of three processes
        in backward.
        At this point, $b_5$, $b_6$, and $b_7$ are
        possible by using the rule backward.
        Since $\mathsf{write}(b_5)=\{\texttt{x}\}$ and both $\mathsf{read}(b_6)$
        and $\mathsf{read}(b_7)$ contain $x$, the order between
        $b_5$\vspace*{0.1cm}
    \end{minipage}
    \mbox{} and $b_6$, and the order between $b_5$ and $b_6$ affect
    the causality. We say $b_i$ {\em conflicts} with $b_j$ where
    $i\not=j$ if
    $\mathsf{read}(b_i)\cap\mathsf{write}(b_j)\not=\varnothing$
    or $\mathsf{read}(b_j)\cap\mathsf{write}(b_i)\not=\varnothing$.
    Since $b_6$ and $b_7$ do not conflict with each other,
    the order between $b_6$ and $b_7$ does not affect the causality.
    Thus, for the forward execution in table~\ref{tbl:exec}\;(a),
    the reversed execution $b_3b_2b_3b_6b_7b_4b_2b_1$ reaches
    $\rho_0$ as a legitimate reversed computation.

    \subsection{Annotation DAG}

    We shall present a data structure called `annotation DAG' (Directed Acyclic Graph) that keeps the conflicting
    information in forward execution and controls the backward execution by
    matching the causality, observing the memory $Wt$ updated by reading
    the memory $Rd$.

\begin{dfn}
An annotation DAG is $A=(V,E_R,E_W)$ satisfying the following conditions:
\begin{enumerate}
\item\label{ADAGcond2} $V\subseteq(\mathsf{PID}\times \mathbb{N})\cup\{\bot\}$ where
$\mathbb{N}$ is the set of natural numbers,
$\bot\in V$, and
if $(p,n)\in V$ then for all $n'\leq n$,
$(p,n')\in V$;
\item\label{ADAGcond3} $E_R,E_W\subseteq V\times\mathcal{R}\times V$ where
$(v',r,v),(v'',r,v)\in E_R\cup E_W$ implies $v'=v''$;
\item\label{ADAGcond1} $E_R\cap E_W=\varnothing$ and
$(V,E_R\uplus E_W)$ is a DAG with the finite set of nodes $V$;
\item\label{ADAGcond4} $(v',r,v)\in E_W$
and $v'\not=\bot$ imply $(v'',r,v')\in E_W$; and
\item\label{ADAGcond5} $(v,r,v'),(v,r,v'')\in E_W$ implies $v'=v''$
\end{enumerate}\vspace*{-0.05cm}
$\mathcal{A}$ is the set of all annotation DAGs, and
$A_{init}$ is $(\{\bot\},\varnothing,\varnothing)$.
\end{dfn}

    We write $v\stackrel{r}{\rightarrow}v'$ for $(v,r,v')\in E_W$
    and $v\stackrel{r}{\dashrightarrow}v'$ for $(v,r,v')\in E_R$.
    Condition~\ref{ADAGcond5} with conditions~\ref{ADAGcond1} and \ref{ADAGcond3} ensures that when
    $v'\tran{r}v$, there is a unique sequence of $E_W$ with the label of $r$ from $\bot$ to $v$:
    $\bot\stackrel{r}{\rightarrow}v_1\stackrel{r}{\rightarrow}\cdots\stackrel{r}{\rightarrow}v_n=v$.
    $\mathsf{last}(r,E_W)$ denotes the last node $v$ of such sequence.
    When $\mathsf{last}(r,E_W)=v\neq \bot$, $v'\tran{r}v$ for a unique $v'$ and
    $v\ntran{r}v''$ for all $v''$.
    $\mathsf{last}(r,\varnothing)=\bot$ for all $r\in\mathcal{R}$.
    Since $V$ is finite, for $(p,n)\in V$
    there is the maximum number for process $p$ if such
    $(p,n)$ exists. Given $V\subseteq\mathsf{PID}\times\mathbb{N}\cup\{\bot\}$, we write
    $\mathsf{max}_p(V)$ for $max_{(p,n)\in V}\;n$ for some $(p,n)\in V$.
    $\mathsf{max}_p(V)=-1$ when $(p,n)\notin V$ for all $n$.

    \begin{dfn}
        For $A_1,A_2\in\mathcal{A}$,
        $A_1=(V_1,{E_R}_1,{E_W}_1)\both{p,Rd,Wt}{ann}A_2=(V_2,{E_R}_2,{E_W}_2)$ if
        \begin{enumerate}
            \item $V_2=V_1\cup\{v\}$;
            \item ${E_R}_2={E_R}_1\cup\{\mathsf{newedge}(r,{E_W}_1,v)\;|\;r\in Rd-Wt\}$;and
            \item ${E_W}_2={E_W}_1\cup\{\mathsf{newedge}(r,{E_W}_1,v)\;|\;r\in Wt\}$
        \end{enumerate}
        where $v=(p,\mathsf{max}_p(V_1)+1)$ and $\mathsf{newedge}(r,E_W,v)=(\mathsf{last}(r,E_W),r,v)$.
    \end{dfn}
    Given $O\subseteq\mathsf{PID}\times 2^\mathcal{R}\times
    2^\mathcal{R}$,
    $A\both{  O  }{ann}^+ A'$ when for some $(p_i,Rd_i,Wt_i)\in O$,
    there exists a sequence of $A_i$ such that
    $A_0\both{p_1,Rd_1,Wt_1}{ann}A_1\cdots\both{p_n,Rd_n,Wt_n}{ann}A_n=A'$.
    We write $A\both{  O  }{ann}^\ast A'$ if $A=A'$ or $A\both{  O  }{ann}^+ A'$.

    Intuitively, the forward (left-to-right) relation
    of $A_1=(V_1,E_{R1},{E_W}_1)\both{p,Rd,Wt}{ann}A_2$
    is to add a fresh node $v$ and edges
    $\mathsf{newedge}(r,{E_W}_1,v)$ for $r\in Rd\cup Wt$ to $A_1$
    for process $p$ to execute forwards
    a basic block $b\in\mathsf{PB}(p)$
    with $\mathsf{read}(b) = Rd$ and $\mathsf{write}(b) = Wt$.
    $v$ presents the new causality created by executing
    process $p$ in the forward direction.
    The new node has the causality
    to update $Wt$ from $(p,\mathsf{max}_p(V_1))$ that presents
    the newest causality in $A_1$ by $p$ in forward. To update
    the causality, the edges $\mathsf{newedge}(r,{E_W}_1,v)$ for $r\in Wt$
    are added to ${E_W}_1$. The edges $\mathsf{newedge}(r,{E_W}_1,v)$ for $r\in Rd-Wt$
    from the newest causality for $r$ at that moment
    are added to ${E_R}_1$ to show that the update for $v$ depends on such $r$.

    The backward (right-to-left) relation is to remove
    a node and edges from $A_2$. The node $v$ to be
    removed has to be the newest causality of a process
    and does not depend on other causalities.
    It is shown that such a node always exists in
    an annotation DAG in $\{A\;|\;A_{init}\both{O}{ann}^+A\}$ as below.

    \begin{prop}
        For $(V,E_R,E_W)\in\{A\;|\;A_{init}\both{\quad}{ann}^+A\}$,
        there exists a node $v\in V$ such that
        $v'\dtran{r}v$ implies
        $v'=\mathsf{last}(r,E_W)$;
        and no outgoing edge from $v$.
    \end{prop}

    Moreover, $\mathcal{A}^O_{comp}=\{A\;|\;A_{init}\both{O}{ann}^\ast A\}$
    is closed by $\both{p,Rd,Wt}{ann}$ where $(p,Rd,Wt)\in O$. Obviously, $A\in\mathcal{A}_{comp}$
    implies $A'\in\mathcal{A}^O_{comp}$ when $A\both{p,Rd,Wt}{ann}A'$ for some $(p,Rd,Wt)\in O$
    by definition.

    \begin{prop}
        \label{annotation-closed}
        For $A\in\{A'\;|\;A_{init}\both{O}{ann}^+ A'\}$,
        there exists $A''\in\mathcal{A}^O_{comp}$ such that
        $A''\both{p,Rd,Wt}{ann}A$ with
        $(p,Rd,Wt)$
        $\in O$.
    \end{prop}

    \subsection{Operational semantics controlled by Annotation DAG}

    \begin{dfn}
        The {\em operational semantics controlled by annotation DAG} over program configurations
        $(C, A) \xrightleftharpoons{p,Rd,Wt} (C', A')$
        is defined by:

        \begin{center}
            \begin{prooftree}
                \AxiomC{$C \both{p,Rd,Wt}{{ prog }} C'$}
                \AxiomC{$A \both{p,Rd,Wt}{{ ann }} A'$}
                \RightLabel{\bf ProgAnn}
                \BinaryInfC{$(C, A) \xrightleftharpoons{p,Rd,Wt} (C', A')$}
            \end{prooftree}
        \end{center}
        where $p\in\mathsf{PID}$ and $Rd,Wt\subseteq\mathcal{R}$.

        The {\em program computation with annotation} is a sequence of
        $(C_i,A_i)\both{p_i,Rd_i,Wt_i}{\quad}(C_{i+1},A_{i+1})$ ($i\geq 0$)
        beginning with $(C_0,A_0)=(C_{init},A_{init})$.

    \end{dfn}

    We illustrate the behavior controlled by the annotation DAG
    for the simple example of the previous section. 
    Starting from the initial configuration,$(C_0,A_0)=
    ((Pg,\rho_0,\sigma_0,[(\varepsilon\mapsto (\texttt{main},\texttt{begin}))]),
    (\{\bot\},\varnothing,\\ \varnothing))$, it ends up with
    $(C_8,A_8)=(P_g,[x,y,z\mapsto 2,1,1],\sigma_0,
    [\varepsilon\mapsto(\texttt{main},\texttt{end})])$.

\begin{figure}[H]
\centering

\subfloat[][$A_3$]{\includegraphics[scale=0.8]{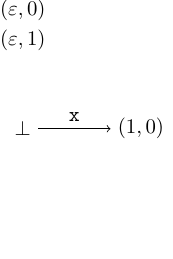}}\label{fig:A3}\quad
\subfloat[][$A_4$]{\includegraphics[scale=0.8]{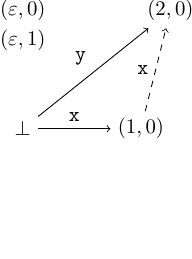}}\label{fig:A4}\quad
\subfloat[][$A_5$]{\includegraphics[scale=0.8]{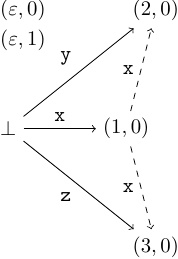}}\label{fig:annon3}\quad
\subfloat[][$A_6$]{\includegraphics[scale=0.8]{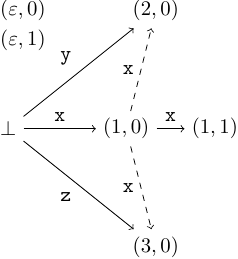}}\label{fig:annot2}\quad
\subfloat[][$A_8$]{\includegraphics[scale=0.8]{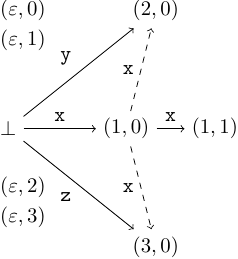}}\label{fig:annot1}

\caption{Annotation DAGs along with forward execution}\label{fig:annot}
\end{figure}

    \paragraph{Forward accumulation of causality}
     We present the construction of
    annotation DAGs as follows:

    \begin{trivlist}
        \item[(F1)] After process $\varepsilon$ executes $b_1$ and $b_2$,
        $A_2=(\{\bot,(\varepsilon,0),(\varepsilon,1)\},\varnothing,\varnothing)$;
        \item[(F2)] The call statement in $b_2$ forks three subprocesses.
        Then, process $1$ executes $b_4$, $(1,0)$ is added to $V$ and
        $\bot\tran{\texttt{x}}(1,0)$ is added since
        $\mathsf{read}(b_4)=\mathsf{write}(b_4)=\{\texttt{x}\}$ to make $A_3$, meaning $\texttt{x}$ is updated by the initial $\texttt{x}$,
        and the store is updated as $[\texttt{x},\texttt{y},\texttt{z}\mapsto 1,0,0]$.
        \item[(F3)] Next, process $2$ executes $b_6$ where $\mathsf{read}(b_6)=\{\texttt{x},\texttt{y}\}$
        and $\mathsf{write}(b_6)=\{\texttt{y}\}$.  $\both{2,\{\texttt{x},\texttt{y}\},\{\texttt{y}\}}{ann}$ adds
        a fresh node $(2,0)$, $\bot\tran{\texttt{y}}(2,0)$, and
        $(1,0)\dtran{\texttt{x}}(2,0)$. The causality of $(2,0)$ means
        $y$ is updated by the initial $\texttt{y}$ and $\texttt{x}$ of $(1,0)$ to make $A_4$.
        \item[(F4)] Then, process $3$ executes $b_7$ where $\mathsf{read}(b_7)=\{\texttt{x},\texttt{z}\}$
        and $\mathsf{write}(b_7)=\{\texttt{z}\}$.  $\both{3,\{\texttt{x},\texttt{z}\},\{\texttt{z}\}}{ann}$ adds
        $(3,0)$, $\bot\tran{\texttt{z}}(3,0)$, and $(1,0)\dtran{\texttt{x}}(3,0)$,
        to make $A_5$ shown in figure~\ref{fig:annot}\,(a), meaning the causality at $(3,0)$ to update the
        initial $\texttt{z}$ using the initial $\texttt{z}$ and $\texttt{x}$ of $(1,0)$.
        \item[(F5)] At last, process $1$ executes $b_5$
        where $\mathsf{read}(b_5)=\mathsf{write}(b_5)=\{\texttt{x}\}$.   $\both{1,\{\texttt{x}\},\{\texttt{x}\}}{ann}$
        just adds $(1,1)$ and $(1,0)\tran{\texttt{x}}(1,1)$
        to form $A_6$ shown in figure~\ref{fig:annot}\,(b), meaning $\texttt{x}$ is updated
        by $\texttt{x}$ of $(1,0)$.
        \item[(F6)] No more causality is
        created after merging the subprocesses. Just the relation adds $(\varepsilon,2)$ and $(\varepsilon,3)$
        with no edges to form $A_8$ shown in figure~\ref{fig:annot}\,(c).
    \end{trivlist}

    \paragraph{Backward rollback of causality}
    The following is the summary of the corresponding
    backward execution.
    \begin{trivlist}
        \item[(B1)]
        The removable nodes
        of $A_8$ are $\{(\varepsilon,3),(1,1)\}$. Here, $C_8$ specifies
        $\varepsilon$ to remove $(\varepsilon,3)$ followed by removing
        $(\varepsilon,2)$ back to $(C_6,A_6)$, where
        $C_6=(Pg,[\texttt{x},\texttt{y},\texttt{z}\mapsto 2,1,1],\sigma_0,[\varepsilon\mapsto
        (\texttt{l2},\texttt{run}),1\mapsto(\texttt{sub0},\texttt{end}),
        2 \\ \mapsto(\texttt{sub1},\texttt{end}),3\mapsto(\texttt{sub2},\texttt{end})])$
        \item[(B2)]
        $C_6$ may reverse any subprocess, but $A_6$ allows only
        to remove $(1,1)$ by $\both{p,Rd,Wt}{ann}$ to obtain $A_5$.
        \item[(B3)]
        After removing $(1,1)$ and $(1,0)\tran{\texttt{x}}(1,1)$ from $A_6$,
        we obtain $A_5$ whose removable nodes are $(2,0)$ and $(3,0)$.
        $(1,0)$ is not removable since $(1,0)$ has two outgoing edges,
        although $(1,0)=\mathsf{last}(\texttt{x},E_W)$.
        \item[(B4)]
        $C_5$ may reverse either process $2$ or process $3$, and let
        process $2$ reverse to become $C'_4$. Then, remove $\bot\tran{y}(2,0)$ and
        $(1,0)\dtran{x}(2,0)$ to obtain $A'_4$ and $[x,y,z\mapsto 1,0,1]$
        as the store $\rho$. Note that  $(C'_4,A'_4)$ did not appear in the forward execution.
        \item[(B5)] From $(C'_4,A'_4)$, process $3$ is reversed
        to remove $(3,0)$, $\bot\tran{\texttt{z}}(3,0)$, and
        $(1,0)\dtran{\texttt{x}}(3,0)$ to obtain $A_3$
        and $[\texttt{x},\texttt{y},\texttt{z}\mapsto 1,0,0]$.
        \item[(B6)] Then, process $1$ is reversed by removing
        $(1,0)$ and $\bot\tran{\texttt{x}}(1,0)$ to obtain
        $A_2=(\{\bot,(\varepsilon,0),(\varepsilon,1)\},\varnothing, \\ \varnothing)$.
        \item[(B7)] At last, process $\varepsilon$ reverses $b_2$ and $b_1$ to obtain
        $(C_{init},A_{init})$.
    \end{trivlist}

    \begin{wrapfigure}[11]{r}{10cm}\vspace*{-0.7cm}
        \centering
        \includegraphics[scale=0.76]{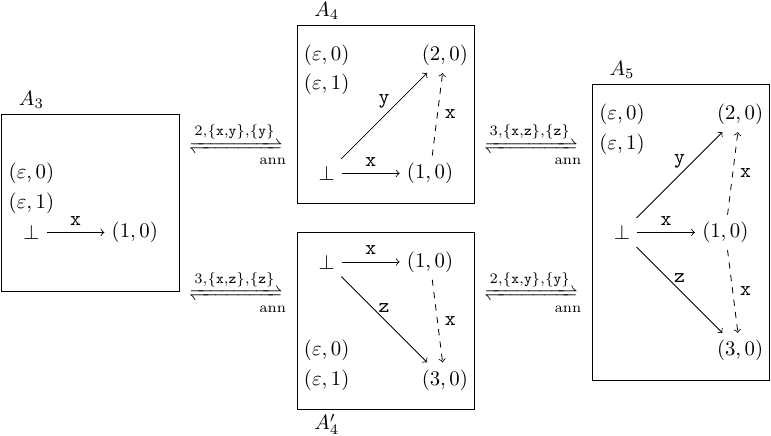}
        \caption{Annotation DAGs in backward execution}
    \end{wrapfigure}

    In (B4) step, there are two possibilities of reversing process 3 or process 2.
    In the above, $A_5$ is reversed by process 2 to $A'_4$ followed by process 3.

    For a CRIL program $Pg$, let $B$ be the basic blocks in $Pg$.
    Let $O=\mathsf{PID}\times\bigcup_{b\in B}\mathsf{read}(b)
    \times\bigcup_{b\in B}\mathsf{write}(b)$.
    Proposition \ref{annotation-closed} ensures there is always
    a removable node along with removable edges.

    \vspace*{0.3cm}

    \subsection{Properties for reversibility}

    We show that the operational semantics controlled by
    annotation DAG has proper properties
    for reversibility.
    We focus on the following two properties that are considered
    fundamental properties for reversibility \cite{DBLP:conf/fossacs/LanesePU20}.

    \begin{description}
        \item [{Causal Safety (CS):}]
        An action can not be reversed until any actions caused by it have been reversed.

        \item [{Causal Liveness (CL):}]
        We should allow actions to reverse in any order compatible with
        Causal Safety, not necessarily the exact inverse of the forward order.
    \end{description}

    \cite{DBLP:conf/fossacs/LanesePU20} shows that those properties
    hold in an LTSI (LTS with Independence) provided that a small number
    of axioms are valid in the LTSI.
    We shall follow this approach by defining LTS from $\both{p,Rd,Wt}{\quad}$
    and add the independence relation
    to have the LTSI for the CRIL behavior.
    We will then show that the axioms for {\bf CS}
    and {\bf CL} hold.

    \begin{dfn}
        $(\mathcal{C}\times\mathcal{A},\mathsf{Lab},\rightharpoonup)$ is the {\em forward
        LTS for CRIL} where:
        \begin{itemize}
            \item $\mathsf{Lab}=\mathsf{PID}\times 2^{\mathcal{R}}\times 2^{\mathcal{R}}$; and
            \item $(C,A)\xrightharpoonup{(p,Rd,Wt)}(C',A')$ if
            $(C,A)\both{p,Rd,Wt}{\quad}(C',A')$
        \end{itemize}
    \end{dfn}

    \begin{dfn}
        The (combined) LTS for CRIL is $(\mathcal{C}\times\mathcal{A},\mathsf{Lab}\uplus\underline{\mathsf{Lab}},\rightarrow)$
        where:
        \begin{itemize}
            \item $\underline{\mathsf{Lab}}=\{\underline{(p,Rd,Wt)}\;|\;(p,Rd,Wt)\in \mathsf{Lab}\}$; and
            \item For $a\in \mathsf{Lab}$, $(C,A)\tran{a}(C',A')$ iff $(C,A)\xrightharpoonup{a}(C',A')$,
            and $(C,A)\tran{\underline{a}}(C',A')$ iff $(C',A')\tran{a}(C,A)$.
        \end{itemize}
    \end{dfn}

    $\Lab\uplus\ULab$ is ranged over by $\alpha,\beta,\cdots$,
    and $\Lab$ by $a,b,\cdots$.  $\mathsf{und}:\Lab\uplus\ULab\rightarrow\Lab$ where
    $\und{a}=a$ and $\und{\underline{a}}=a$.  $\underline{\underline{a}}=a$.
    Given $t:P\tran{a}Q$, $\underline{t}$ is for $Q\tran{\underline{a}}P$.

    For CRIL, the independence of transitions is defined as the
    independent memory update among concurrent processes.
    The processes running concurrently are not in the subprocess
    relation.
    Note that as pid $p\cdot 1$, $p\cdot 2$, $\cdots$ are assigned to the subprocesses of
    the process with pid of $p$.  The process with the pid of $p$ is concurrent to the
    process with the pid of $q$ if $p\not\preceq q$ and $q\not\preceq p$.  Hence, we
    give the dependence relation for labels as follows.

    \begin{dfn}
        For $\alpha,\beta\in\Lab\uplus\ULab$ such that
        $\und{\alpha}=(p_1,Rd_1,Wt_1)$ and $\und{\beta}=(p_2,Rd_2,Wt_2)$,
        $\alpha\ \idplab\ \beta$ iff
        \[ p_1\not\preceq p_2\ \wedge\ p_2\not\preceq p_1\
        \wedge\ Rd_1\cap Wt_2=\varnothing\ \wedge Rd_2\cap Wt_1=\varnothing\]
    \end{dfn}

    The independence of transitions in LTS is defined as the transitions with
    independent labels.  We define the Labeled Transition System with Independent
    transitions as the operational semantics of CRIL.

    \begin{dfn}
        For $t:(C_1,A_1)\xrightarrow{\alpha}(C_1',A_1')$
        and $u:(C_2,A_2)\xrightarrow{\beta}(C_2',A_2')$ in
        the combined LTS for CRIL, $t$ and $u$ are
            {\em independent} of each other, written as $t\ \idp\ u$ if
        $\alpha\ \idplab\ \beta$.

        $(\mathcal{C}\times\mathcal{A},\Lab\uplus\ULab,\rightarrow,\idp)$ is the
            {\em LTS of CRIL with independence}.
    \end{dfn}

    In the sequel, we write `$\ltsi$' for the LTS of CRIL with independence.

    \subsubsection{Basic properties for reversibility}
    We take the axiomatic approach of \cite{DBLP:conf/fossacs/LanesePU20}, where the combination of the
    basic properties gives the proper reversibility.   The first step is to
    show that the $\ltsi$ is {\em pre-reversible}.
    For this purpose, we show $\ltsi$ satisfies the following axioms:
    ``{\bf Square Property (SP)}'',
    ``{\bf Backward Transitions are Independent (BTI)}'',
    ``{\bf Well-Foundedness (WF)}'', and
    ``{\bf Coinitial Propagation of Independence (CPI)}''.

    \paragraph{Square Property(SP)}

    For $a\in\mathsf{Lab}$, when $C\both{\ \ a\ \ }{\mathrm{{\ prog\ }}}C'$, we write
    $C\tran{a}_{\mathrm{\ prog\ }}C'$ and $C'\tran{\underline{a}}_{\mathrm{\ prog\ }}C$.
    Similarly, when $A\both{\ \ a\ \ }{\mathrm{\ ann\ }}A'$, we write
    $A\tran{a}_{\mathrm{\ ann\ }}A'$ and $A'\tran{\underline{a}}_{\mathrm{\ ann\ }}A$.

    By the definition of the independence transitions, the square property of
    the $\tran{\alpha}_{\mathrm{\ prog\ }}$ is immediately shown.

    \begin{prop}
        \label{prop:par-sp}
        Suppose $C \xrightarrow{\alpha}_{\mathrm{{ prog }}} C'$, $C \xrightarrow{\beta}_{\mathrm{{ prog }}} C''$,
        and $\alpha \idplab \beta$.
        Then there are the cofinal transitions $C' \xrightarrow{\beta}_{\mathrm{{ prog }}} C'''$ and
        $C'' \xrightarrow{\alpha}_{\mathrm{{ prog }}} C'''$.
    \end{prop}

    For annotation DAGs,  we need to trace the difference of nodes and edges
    added or deleted by $\anntran{\alpha}$ to show the square property.
    We use the following notation to present differences in annotation DAGs:
    \begin{trivlist}
        \item
        For $o : (V,E_R,E_W)\anntran{\alpha}(V',E'_R,E'_W)$,
        $\mathsf{diff}(o)=
        \begin{cases}
        (V' - V,E'_R - E_R,E'_W - E_W) & \mbox{if }\alpha\in\Lab,\\
        (V - V',E_R - E'_R,E_W - E'_W) & \mbox{if }\alpha\in\ULab
        \end{cases}
        $
        \item
        $(V,E_R,E_W) \odot^\alpha (\Delta V,{\Delta E_R},{\Delta E_W})=
        \begin{cases}
        (V\cup \Delta V,E_R\cup {\Delta E_R},E_W\cup {\Delta E_W}) &
        \mbox{if }\alpha\in\Lab,\\
        (V- \Delta V,E_R- {\Delta E_R},E_W- {\Delta E_W}) &
        \mbox{if }\alpha\in\ULab
        \end{cases}$
    \end{trivlist}

    \begin{prop}
        \label{prop:diff-disjoint}
        Let $\mathsf{diff}(A \anntran{\alpha} A') = (\Delta V^\alpha,\Delta E_R^\alpha,\Delta E_W^\alpha)$
        and $\mathsf{diff}(A \anntran{\beta} A'') = (\Delta V^\beta,\Delta E_R^\beta,\Delta E_W^\beta)$
        with $\alpha\idp_\mathrm{lab}\beta$.
        Then, $\Delta V^\alpha\cap \Delta V^\beta=\Delta E_{R}^\alpha\cap \Delta E^\beta_{R}=\Delta E^\alpha_{W}\cap \Delta E^\beta_{W}=\varnothing$.
    \end{prop}
    \begin{proof}
        For some $v_\alpha$ and $v_\beta$,
        $\Delta V^\alpha = \{v_\alpha\}$ and $\Delta V^\beta = \{v_\beta\}$.
        $\alpha\idp_\mathrm{lab}\beta$ implies $v_\alpha \neq v_\beta$.
        All the edges of $\Delta E_{R}^\alpha\uplus \Delta E^\alpha_{W}$ come into $v_\alpha$
        and all the edges of $\Delta E_{R}^\beta\uplus \Delta E^\beta_{W}$ come into $v_\beta$.
        Therefore, $\Delta V^\alpha\cap \Delta V^\beta=\Delta E_{R}^\alpha\cap \Delta E^\beta_{R}=\Delta E^\alpha_{W}\cap \Delta E^\beta_{W}=\varnothing$.
    \end{proof}

    \begin{prop}
        \label{prop:diff-preserve-f}
        Suppose $A \anntran{a} A'$
        and $A \anntran{\beta} A''$
        with $a\idp_\mathrm{lab}\beta$.
        Then
        there is $A'''$ such that
        $A'' \anntran{a} A'''$
        and $\mathsf{diff}(A \anntran{a} A')
        = \mathsf{diff}(A'' \anntran{a} A''')$.
    \end{prop}
    \begin{proof}
        Assume $A=(V,E_R,E_W)$, $A''=(V'',E''_R,E''_W)$,
        and $a=(p_a,Rd_a,Wt_a)$.
        $A'' \anntran{a} A'''$ for some $A'''$ since $a\in\Lab$.
        $a\idp_\mathrm{lab}\beta$ implies that
        $\mathsf{max}_{p_a}(A) = \mathsf{max}_{p_a}(A'')$ and
        $\mathsf{last}(r,E_W) = \mathsf{last}(r,E''_W)$
        for $r\in Rd_a$.
        Therefore, $\mathsf{diff}(A \anntran{a} A')
        = \mathsf{diff}(A'' \anntran{a} A''')$.
    \end{proof}

    \begin{prop}
        \label{prop:diff-preserve-b}
        Suppose $A \anntran{\underline{a}} A'$
        and $A \anntran{\beta} A''$
        with $\underline{a}\idp_\mathrm{lab}\beta$.
        Then
        there is $A'''$ such that
        $A'' \anntran{\underline{a}} A'''$
        and $\mathsf{diff}(A \anntran{\underline{a}} A')
        = \mathsf{diff}(A'' \anntran{\underline{a}} A''')$.
    \end{prop}
    \begin{proof}
        Assume $\mathsf{diff}(A \anntran{\beta} A'')
        = (\Delta V^\beta,\Delta E_R^\beta, \Delta E_W^\beta)$,
        and $a=(p_a,Rd_a,Wt_a)$.
        Let $v = (p_a, \mathsf{max}_{p_a}(V))$.
        $\underline{a}\idp_\mathrm{lab}\beta$ implies that
        no edges in $\Delta E^\beta_R \uplus \Delta E^\beta_W$ go out from $v$
        and $v'$ such that $v' \dtran{r} v$ in $A$.
        Therefore, $A'' \anntran{\underline{a}} A'''$ for some $A'''$.
        $\underline{a}\idp_\mathrm{lab}\beta$ and $\underline{a}\in\ULab$ derive
        $\mathsf{diff}(A \anntran{\underline{a}} A')
        = \mathsf{diff}(A'' \anntran{\underline{a}} A''')$.
    \end{proof}

    \begin{prop}
        \label{prop:diff-preserve}
        Suppose $A \anntran{\alpha} A'$
        and $A \anntran{\beta} A''$
        with $\alpha\idp_\mathrm{lab}\beta$.
        Then $A'' \anntran{\alpha} A'''$,
        where $A''' = A'' \odot^\alpha \mathsf{diff}(A \anntran{\alpha} A')$.
    \end{prop}
    \begin{proof}
        Proposition~\ref{prop:diff-preserve-f} and~\ref{prop:diff-preserve-b}
        derive $A'' \anntran{\alpha} A'''$.
    \end{proof}

    \begin{prop}
        \label{prop:ann-sp}
        Suppose $A \xrightarrow{\alpha}_{\mathrm{ann}} A'$,
        $A \xrightarrow{\beta}_{\mathrm{ann}} A''$,
        and $\alpha \idp_\mathrm{lab} \beta$.
        Then there are the cofinal transitions
        $A' \xrightarrow{\beta}_{\mathrm{ann}} A'''$ and
        $A'' \xrightarrow{\alpha}_{\mathrm{ann}} A'''$.
    \end{prop}
    \begin{proof}
        By proposition~\ref{prop:diff-disjoint}, $\mathsf{diff}(A\anntran{\alpha}A')$ and
        $\mathsf{diff}(A\anntran{\beta}A'')$ are disjoint if $\alpha\idp_\mathrm{lab}\beta$.  Hence, the order of
        addition and deletion to/from $A$ does not affect the result.  Therefore,
        $(A \odot^{\alpha}\mathsf{diff}(A\anntran{\alpha} A'))
        \odot^{\beta}\mathsf{diff}(A\anntran{\beta} A'')=
        (A\odot^{\beta}\mathsf{diff}(A \anntran{\beta} A''))
        \odot^{\alpha}\mathsf{diff}(A \anntran{\alpha} A')=A'''$.
        By proposition~\ref{prop:diff-preserve},
        we have $A\anntran{\alpha}A'\anntran{\beta}A'''$ and $A\anntran{\beta}A''\anntran{\alpha}A'''$
        hold for such $A'''$.
    \end{proof}

    Combining proposition~\ref{prop:par-sp} with proposition~\ref{prop:ann-sp}
    by {\bf ProgAnn}, the square property holds.
    \begin{lemm}[Square Property]
        \label{prop:sp}
        Whenever
        $t : { (C_P,A_P) } \xrightarrow{\alpha} { (C_Q,A_Q) }$,
        $u : { (C_P,A_P) } \xrightarrow{\beta} { (C_R,A_R) }$,
        and $t \idp u$,
        then there are cofinal transitions
        $u' : { (C_Q,A_Q) } \xrightarrow{\beta} { (C_S,A_S) }$,
        and $t' : { (C_R,A_R) } \xrightarrow{\alpha} { (C_S,A_S) }$.
    \end{lemm}

    \paragraph{Backward Transitions are Independent (BTI)}

    {\bf BTI} is useful for reversibility because an available backward
    transition does not depend on any other backward transition.   In CRIL,
    a label of $\ltsi$ gives the information to establish BTI.

    \begin{lemm}[Backward Transitions are Independent]
        \label{prop:bti}
        Whenever $t : { (C_P,A_P) } \xrightarrow{\underline{a}} { (C_Q,A_Q) }$,
        $u : { (C_P,A_P) } \xrightarrow{\underline{b}} { (C_R,A_R) }$,
        and $t \neq u$,
        then $t \idp u$.
    \end{lemm}
    \begin{proof}
        Assume $A_P = (V,E_R,E_W)$,
        $a = (p_a,Rd_a,Wt_a)$, and $b = (p_b,Rd_b,Wt_b)$.
        Let $v_a = (p_a, \mathsf{max}_{p_a}(V))$
        and $v_b = (p_b, \mathsf{max}_{p_b}(V))$.

        Assume $p_a \preceq p_b$.
        Then $p_a = p_b$ holds from the operational semantics.
        $p_a = p_b$ derives $t = u$, which contradicts $t \neq u$.
        Therefore, $p_a \not\preceq p_b$ holds.
        Similarly, $p_b \not\preceq p_a$ also holds.

        Assume $Rd_a \cap Wt_b \neq \varnothing$.
        There exists $ r  \in Rd_a \cap Wt_b$.
        If $ r  \in Wt_a$, then $\mathsf{last}(r,E_W) = v_a$ and
        $\mathsf{last}(r,E_W) = v_b$.
        Therefore $p_a = p_b$, however it contradicts $p_a \not\preceq p_b$.
        If $r \not\in Wt_a$, then $\mathsf{last}(r,E_W)\dtran{r}v_a \in E_R$.
        $r \in Wt_b$ derives $\mathsf{last}(r,E_W) = v_b$.
        Therefore $v_b\dtran{r}v_a\in E_R$, however it contradicts
        that no edges go out from $v_b$ derived from $u$.
        Therefore $Rd_a \cap Wt_b = \varnothing$.
        Similarly, $Rd_b \cap Wt_a = \varnothing$ also holds.
    \end{proof}

    \paragraph{Well-Foundedness (WF)}
    For a backward transition $(C,A)\tran{\underline{a}}(C',A')$, the number of nodes of $A'$
    is strictly less than that of $A$.
    Since the number of vertices of annotation DAG is finite,
    it is not possible to remove vertices infinitely.

    \paragraph{Coinitial Propagation of Independence (CPI)}

    Given a commuting square with independence at one corner, CPI allows us
    to deduce independence between coinitial transitions at the other three corners.

    \begin{lemm}[Coinitial Propagation of Independence]
        \label{prop:cpi}
        Suppose
        $t : { (C_P,A_P) } \xrightarrow{\alpha} { (C_Q,A_Q) }$,
        $u : { (C_P,A_P) } \xrightarrow{\beta} { (C_R,A_R) }$,
        $u': { (C_Q,A_Q) } \xrightarrow{\beta} { (C_S,A_S) }$,
        $t': { (C_R,A_R) } \xrightarrow{\alpha} { (C_S,A_S) }$,
        and $t \idp u$.
        Then $u' \idp \underline{t}$.
    \end{lemm}

    \begin{proof}
        $t\idp u$ implies $\alpha\idplab \beta$.
        Since $\beta\idplab \underline{\alpha}$,
        $u'\idp\underline{t}$.
    \end{proof}

    \subsubsection{Events}

    The properties above make $\ltsi$ pre-reversible.
    Next, we check if $\ltsi$ can derive
    events for establishing reversibility.
    Following \cite{DBLP:conf/fossacs/LanesePU20},
    events in $\ltsi$ are derived as an equivalence over transitions.

    \begin{dfn}
        Let $\sim$ be the smallest equivalence relation on transitions satisfying:
        if $t : { (C_P,A_P) } \xrightarrow{\alpha} { (C_Q,A_Q) }$,
        $u : { (C_P,A_P) } \xrightarrow{\beta} { (C_R,A_R) }$,
        $u' : { (C_Q,A_Q) } \xrightarrow{\beta} { (C_S,A_S) }$,
        $t' : { (C_R,A_R) } \xrightarrow{\alpha} { (C_S,A_S) }$,
        and $t \idp u$,
        then $t \sim t'$.
        The equivalence classes of forward transitions $[(C_P,A_P)\tran{a}(C_Q,A_Q)]$,
        are the {\em events}.
        The equivalence classes of backward transitions $[(C_P,A_P)\tran{\underline{a}}(C_Q,A_Q)]$,
        are the {\em reverse events}.
    \end{dfn}

    Given $\gamma=\alpha_1\cdots\alpha_n\in(\Lab\uplus\ULab)^\ast$, a sequence of transitions
    $(C_0,A_0)\tran{\alpha_1}\cdots\tran{\alpha_n}(C_n,A_n)$ is written as $s:(C_0,A_0)\tran{\gamma}_\ast
    (C_n,A_n)$.

    Since the transitions of program configurations in $\ltsi$ $\tran{\alpha}_{\mathrm{prog}}$
    have no control for reversibility,
    events are substantially derived from the operations of annotation DAGs.

    \begin{dfn}
        Let $\sim_{\mathrm{ann}}$ be the smallest equivalence relation over operations of annotation DAGs satisfying:
        if $o_1: A_P \xrightarrow{\alpha}_{\mathrm{ann}} A_Q$,
        $o_2: A_P \xrightarrow{\beta}_{\mathrm{ann}} A_R$,
        $o_2': A_Q  \xrightarrow{\beta}_{\mathrm{ann}} A_S$,
        $o_1': A_R \xrightarrow{\alpha}_{\mathrm{ann}} A_S$,
        and $\alpha \idplab \beta$,
        then $o_1 \sim o'_1$.  $[A\anntran{a}A']_{\mathrm{ann}}$ and $[A\anntran{\underline{a}}A']_{\mathrm{ann}}$ are the forward and
        backward equivalence classes by $\sim_{\mathrm{ann}}$.
    \end{dfn}

    \begin{prop}
        \label{prop:DagCausality}
        For $t : { (C_P,A_P) } \xrightarrow{\alpha} { (C_Q,A_Q) }$ and
        $t' : { (C_R,A_R) } \xrightarrow{\alpha} { (C_S,A_S) }$, the following holds.\\
        $t\sim t'$
        iff $o\sim_{\mathrm{ann}} o'$
        and $\exists \gamma. (C_P,A_P) \tran{\gamma}_\ast (C_R,A_R)$
        where $o: A_P\anntran{\alpha}A_Q$ and $o': A_R\anntran{\alpha}A_S$.
    \end{prop}

    Intuitively, operations for annotation DAGs are independent if they add or remove
    nodes and edges at unrelated places. If $o_1\sim_{\mathrm{ann}}o_2$, then $o_1$ and
    $o_2$ add or remove the same fragment of annotation DAGs to or from the nodes
    of the same causality. In $\ltsi$, the equivalence over operations of annotation DAGs
    is considered as an {\em event}.
    This shows that events for reversibility are consistently defined over
    $\ltsi$, meaning the operational semantics is detailed enough to give
    the {\bf IRE} property as follows, which is necessary for our objectives.

    \paragraph{Independence Respects Events (IRE)}

    \begin{lemm}[Independence Respects Events]
        \label{prop:ire}
        Suppose $t \sim t' \idp u$.
        Then $t \idp u$.
    \end{lemm}

    \begin{proof}
        If $t\sim t'$, $t$ has the same label as $t'$. Then, $t\idp u$.
    \end{proof}

    \subsubsection{Causal Safety and Causal Liveness}

    Let $\sharp(s,[A\tran{a}A']_{\mathrm{ann}})$ be the number of occurrences of transitions $t$ in $s$
    such that $t \in [(C,A)\tran{a}(C',A')]$, minus
    the number of occurrences of transitions $t$ in $s$ such that
    $t \in [(C,A)\tran{\underline{a}}(C',A')]$.

    Using the result of \cite{DBLP:conf/fossacs/LanesePU20}, the properties of {\bf SP}(Lemma~\ref{prop:sp}), {\bf BTI}(Lemma~\ref{prop:bti}),
        {\bf WF}, {\bf CPI}(Lemma~\ref{prop:cpi}), and {\bf IRE} (Lemma~\ref{prop:ire}) make
        {\bf Causal Safety (CS)} and {\bf Causal Liveness (CL)} hold. Due to the fact that the causality is stored in the
    annotation DAGs, the properties can be stated in $\ltsi$ as below.

    \begin{theo}[Causal Safety]
        \label{prop:cs}
        Whenever $(C_P,A_P) \xrightarrow{a} (C_Q,A_Q)$,
        $s : (C_Q,A_Q) \xrightarrow{\gamma}_* (C_R,A_R)$ with
        $\sharp(s, [A_P\tran{a}A_Q]_{\mathrm{ann}}) = 0$,
        and $(C_S,A_S) \xrightarrow{a} (C_R,A_R)$
        then $(C_P,A_P)\tran{a}(C_Q,A_Q) \idp t$ for all $t$ in $s$
        such that $\sharp(s,[A_P\tran{a}A_Q]_{\mathrm{ann}}) > 0$.
    \end{theo}

    \begin{theo}[Causal Liveness]
        \label{prop:cl}
        Whenever ${ (C_P,A_P) } \xrightarrow{a} { (C_Q,A_Q) }$,
        $s : { (C_Q,A_Q) } \xrightarrow{\gamma}_* { (C_R,A_R) }$,
        $\sharp(s, [A_P\tran{a}A_Q]) = 0$,
        and ${ (C_P,A_P) }\tran{a}{ (C_Q,A_Q) } \idp t:(C,A)\tran{b}(C',A')$ for all $t$ in $s$
        such that $\sharp(s, [A\tran{a}A']) > 0$
        with ${ (C_P,A_P) }\tran{a}{ (C_Q,A_Q) } \sim { (C_S,A_S) }\tran{a}{ (C_R,A_R) }$,
        then we have ${ (C_S,A_S) } \xrightarrow{a} { (C_R,A_R) }$
        with ${ (C_P,A_P) }\tran{a}{ (C_Q,A_Q) } \sim { (C_S,A_S) }\tran{a}{ (C_R,A_R) }$.
    \end{theo}

    Based on these properties, $\ltsi$ can be implemented correctly
    with the pointers for processes managed by a process map along with annotation DAGs
    as the operational semantics of CRIL.

    \section{Example: Airline ticketing}

    We show a version of the airline ticketing program \cite{DBLP:series/lncs/HoeyL0UV20} in CRIL in
    figure~\ref{fig:airline}.
    Two agents attempt to
    sell three seats of an airline.
    This program has a data race for variable $\texttt{seats}$ of

    \hspace*{-0.6cm}\begin{minipage}[b]{0.74\textwidth}

        the remaining seats because two agents may check the remaining
        seats simultaneously before making sales.
        Since the data race does
        not always happen, it is useful to roll back to the point where
        checking remaining seats is insufficient.  Here, $\texttt{agent1}$ and
        $\texttt{agent2}$ are used to record the number of tickets sold
        by each agent.

        \begin{figure}[H]
            \centering
            \hspace*{-0.2cm}
            \includegraphics[scale=0.78]{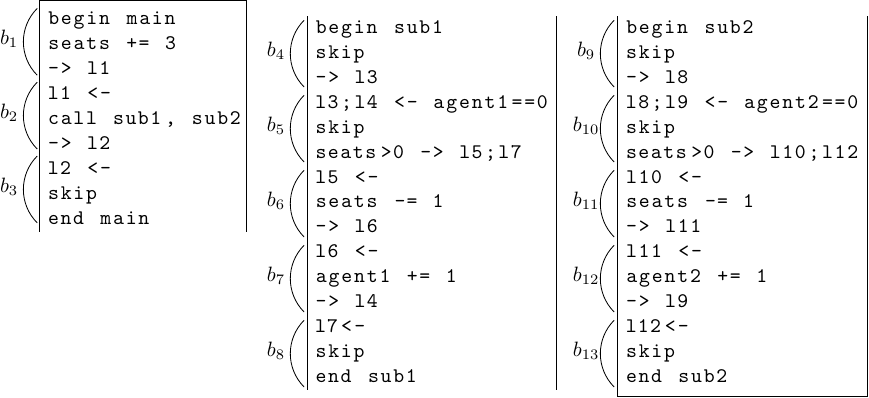}
            \caption{An airline ticketing program in CRIL}
            \label{fig:airline}
        \end{figure}
    \end{minipage}
    \hspace*{-0.15cm}
    \begin{minipage}[b]{0.26\textwidth}
        \begin{table}[H]
            \scriptsize
            \centering
            \hspace*{0.1cm}
            \begin{tabular}{|wc{0.33cm}|wc{0.2cm}|wc{0.3cm}|wc{0.4cm}|wc{0.4cm}|}
                \hline
                \multicolumn{2}{|wc{0.52cm}|}{basic block} & seats & agent1 & agent2 \\\hline
                $(\varepsilon, 0)$ & $b_1$             & 3 & 0 & 0 \\
                $(\varepsilon, 1)$ & $b_2$             & 3 & 0 & 0 \\
                $(1, 0)$           & $b_4$             & 3 & 0 & 0 \\
                $(2, 0)$           & $b_9$             & 3 & 0 & 0 \\
                $(1, 1)$           & $b_5$             & 3 & 0 & 0 \\
                $(1, 2)$           & $b_6$             & 2 & 0 & 0 \\
                $(1, 3)$           & $b_7$             & 2 & 1 & 0 \\
                $(2, 1)$           & $b_{10}$          & 2 & 1 & 0 \\
                $(2, 2)$           & $\mathbf{b_{11}}$ & 1 & 1 & 0 \\
                $(2, 3)$           & $b_{12}$          & 1 & 1 & 1 \\
                $(2, 4)$           & $\mathbf{b_{10}}$ & 1 & 1 & 1 \\
                $(1, 4)$           & $\mathbf{b_5}$ & 1  & 1 & 1 \\
                $(2, 5)$           & $b_{11}$       & 0  & 1 & 1 \\
                $(1, 5)$           & $b_6$          & -1 & 1 & 1 \\
                $(2, 6)$           & $b_{12}$       & -1 & 1 & 2 \\
                $(2, 7)$           & $b_{10}$       & -1 & 1 & 2 \\
                $(2, 8)$           & $b_{13}$       & -1 & 1 & 2 \\
                $(1, 6)$           & $b_7$          & -1 & 2 & 2 \\
                $(1, 7)$           & $b_5$          & -1 & 2 & 2 \\
                $(1, 8)$           & $b_8$          & -1 & 2 & 2 \\
                $(\varepsilon, 2)$ & $b_2$          & -1 & 2 & 2 \\
                $(\varepsilon, 3)$ & $b_3$          & -1 & 2 & 2 \\
                \hline
            \end{tabular}
            \caption{A faulty execution}\label{tab:bug}
        \end{table}
    \end{minipage}

\begin{wrapfigure}[16]{r}{6cm}
        \centering
        \includegraphics[scale=0.75]{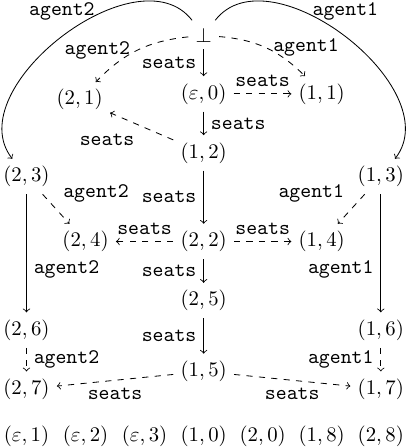}
        \caption{The annotation DAG after the forward execution with the data race}
        \label{fig:bug-annot}
    \end{wrapfigure}

    Table~\ref{tab:bug} shows a forward execution
    that ends $\mathtt{seats}=-1$.
    Figure~\ref{fig:bug-annot} is the annotation DAG
    when terminated at $\texttt{`end\ main'}$ in $b_3$.
    To investigate the cause of the data race, we
    focus on the edges labeled with $\texttt{seats}$.
    The solid edges indicate that
    $\texttt{seats}$ is written in
    $(\varepsilon,0)$, $(1,2)$, $(2,2)$, $(2,5)$,
    and $(1,5)$.
    In particular, $\texttt{seats}$ defined at
    $(2,2)$ is used to update by processes 2 and 3 to cause
    the data race.  (The steps in bold are involved in the problem.)
    To resolve the data race,
    each value of \texttt{seats} should be checked exactly once,
    except for the last value of \texttt{seats}.

    Figure~\ref{fig:airline-vp} shows the airline program where
    $\texttt{sub1}$ and $\texttt{sub2}$ are replaced by those with
    the V-P operations.
    The parameter of the V-P operations works as a semaphore to check and update $\texttt{seats}$
    as a critical region.
    Figure~\ref{fig:debug-annot} is the annotation DAG by the forward execution
    with $\texttt{sub1}$ done first once and then $\texttt{sub2}$ done twice.
    Process 1 executes $b'_5$ setting $\texttt{semaphore}=1$
    at $(1,1)$ first.  ($\texttt{sem}$ is for $\texttt{semaphore}$ in the figure.)
    This prevents process 2 executing $b'_{10}$ at $(2,1)$ since
    $\texttt{semaphore}$ must be 0. Backwards, $b'_{14}$
    and $b'_{15}$ work as $\texttt{V\ semaphore}$.  In the backward execution,
    the order of basic blocks is stored in the annotation DAG.  It works as follows:
    \begin{itemize}
        \item The sequence of $\tran{\texttt{sem}}$ is alternatively from $\texttt{V}$
        and $\texttt{P}$ operations in the forward execution.  $\bot\xrightarrow{\texttt{sem}}(1,1)$
        is by $b'_5$ and $(1,1)\xrightarrow{\texttt{sem}}(1,3)$ by $b'_{14}$, $\cdots$,
        $(1,3)\xrightarrow{\texttt{sem}}(2,1)$ by $b'_{10}$, $(2,1)\xrightarrow{\texttt{sem}}(2,3)$ by $b'_{15}$,$\cdots$.
        \item When $\texttt{seats}=0$, $\texttt{semaphore}$ is released with no operation.\\
        $(2,7)\semtran (1,5)\semtran (1,6)$ by $b'_5$ and $b'_8$ and $(1,6)\semtran (2,9)\semtran (2,10)$
        by $b'_{10}$ and $b'_{13}$.
    \end{itemize}

    \hspace*{-0.7cm}
    \begin{minipage}[b]{0.56\textwidth}
        \begin{itemize}
            \item In backward, $\texttt{sub2}$ is ready since $(2,10)$ is $\mathsf{last}(E_W,\texttt{sem})$.
            \item Then, $\texttt{sub1}$ is done with no operation and $(2,7)$ is $\texttt{P}$ in $\texttt{sub2}$.
            The order of $\texttt{V}$ and $\texttt{P}$ is kept until reaching $\bot$.
        \end{itemize}

        \begin{figure}[H]
            \centering
            \includegraphics[scale=0.8]{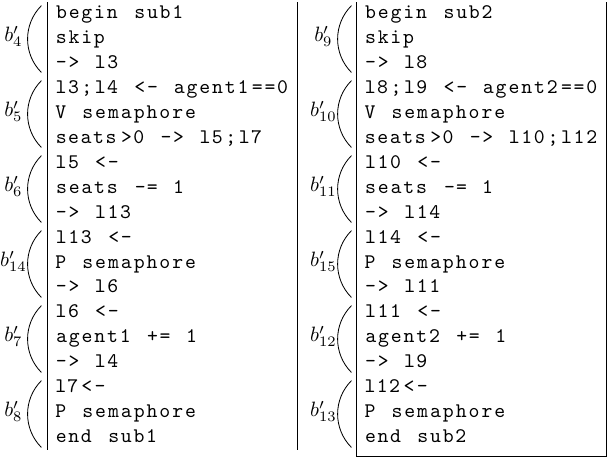}
            \caption{An airline ticketing with semaphore}
            \label{fig:airline-vp}
        \end{figure}
    \end{minipage}
    \begin{minipage}[b]{0.4\textwidth}
        \begin{figure}[H]
            \centering
            \includegraphics[scale=0.75]{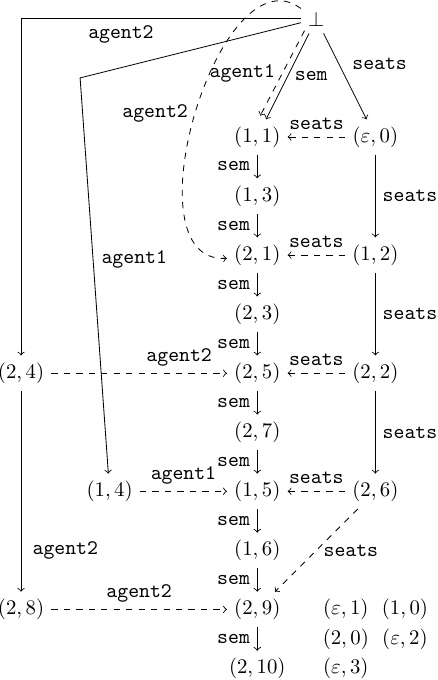}
            \caption{The annotation DAG after the forward execution with semaphore}
            \label{fig:debug-annot}
        \end{figure}
    \end{minipage}

\section{Concluding remarks}

We have proposed CRIL as a reversible concurrent intermediate language.
CRIL is an extension of RIL \cite{DBLP:conf/rc/Mogensen15} to enable running multiple
subroutines as processes running in parallel.
CRIL is intended to be fairly low-level in that each instruction is at a level
similar to the three-address codes to mediate
the translation from a high-level program to a machine-oriented code.
The operational semantics of CRIL defined as $\ltsi$ is shown to have
the properties of Causal Safety and Causal Liveness under the independence of
concurrent processes and shared memory update. By the result of \cite{DBLP:conf/fossacs/LanesePU20},
$\ltsi$ also satisfies other properties: Parabolic lemma, Causal Consistency, Unique
Transition, and Independence of Diamonds.

As related work,
\cite{DBLP:conf/rc/CservenkaGHM18} provides a compiler from  ROOPL++ 
to PISA \cite{DBLP:phd/ndltd/Vieri99}
with no intermediate language, where the translation from an object-oriented source program to the low-level PISA code 
is a big task.
\cite{DBLP:journals/scp/HoeyU22} proposes an annotation to a concurrent imperative
program while executing forward, where the annotation is attached directly to the source program
for reversing the execution.
\cite{DBLP:conf/rc/HoeyU22} investigates its properties of reversibility.  
CRIL uses a similar idea as Hoey's, but CRIL is at a rather lower level 
to provide a finer granularity for 
detailed analysis in translation, such as optimization.  
\cite{DBLP:conf/rc/IkedaY20} presents a collection of 
simple stack machines with a fork and merge
mechanism, where the causality is embedded in the runtime.

For future work, we have focused only on the fundamental properties.  We will investigate further how more properties in reversibility
contribute to behavioral analysis for concurrent programs.  Currently, the dependency of
the heap memory $\texttt{M}$ is treated as one memory resource.
More detailed dependency is necessary for practical use.
Deriving the optimization technique in the front-end part of compilers is
future work via the reversible version of SSA, such as RSSA \cite{DBLP:conf/ershov/Mogensen15} for
concurrent imperative programs.
CRIL is based on the shared memory model.
Incorporating channel-based communications is also future work
to use for the message-passing model like Erlang \cite{DBLP:journals/jlp/LaneseNPV18}.

\paragraph*{Acknowledgement}

We thank Dr. Irek Ulidowski of the University of Leicester for giving valuable suggestions to the draft.
We also thank Prof. Nobuko Yoshida of the University of Oxford, Prof. Hiroyuki
Seki, Prof. Koji Nakazawa, and Prof. Yuichi Kaji of Nagoya University for fruitful discussions.
We thank the anonymous reviewers for providing fruitful comments.
This work is supported by JSPS Kakenhi 21H03415.

    \bibliographystyle{eptcs}
    \bibliography{sos2023-ref}
\end{document}